\documentclass[pra,twocolumn,amsmath,preprintnumbers,amssymb,superscriptaddress,groupedaddress,12pt,nofootinbib]{revtex4}
\usepackage{graphicx}
\usepackage{pdflscape}
\usepackage{verbatim}
 \usepackage{epstopdf}
\usepackage[figuresright]{rotating}
\usepackage{array}
\newcolumntype{P}[1]{>{\raggedright\arraybackslash}p{#1}}
\providecommand{\e}[1]{\ensuremath{\times 10^{#1}}}

 \usepackage{graphics}
\usepackage{amsmath, amsthm, amssymb}

\begin{document}
\title{Optimal pair generation rate for Entanglement-based QKD}
\author{Catherine~Holloway}
\email{c2hollow@iqc.ca}
\affiliation{Institute for Quantum Computing, University of Waterloo, 200 University Ave.~West, Waterloo, Ontario, CA
N2L 3G1}
\author{ John~A.~Doucette} 
\affiliation{David R. Cheriton School of Computer Science, University of Waterloo, 200 University Ave.~West, Waterloo, Ontario, CA
N2L 3G1}

\author{Christopher Erven}
\affiliation{Institute for Quantum Computing, University of Waterloo, 200 University Ave.~West, Waterloo, Ontario, CA
N2L 3G1}

\author{Jean-Philippe Bourgoin}
\affiliation{Institute for Quantum Computing, University of Waterloo, 200 University Ave.~West, Waterloo, Ontario, CA
N2L 3G1}\author{Thomas Jennewein} 
 \affiliation{Institute for Quantum Computing, University of Waterloo, 200 University Ave.~West, Waterloo, Ontario, CA
N2L 3G1}
%\date{\today}

\begin{abstract}
In entanglement-based quantum key distribution (QKD), the generation and detection of multi-photon modes leads to a trade-off between entanglement visibility and two-fold coincidence events when maximizing the secure key rate (SKR). We produce a  predictive model for the optimal two-fold coincidence probability per coincidence window given the channel efficiency and detector dark count rate of a given system. This model is experimentally validated and used in simulations for QKD with satellites as well as optical fibers. 
\end{abstract}
\maketitle

\section*{Introduction}

In entanglement-based QKD a choice must be made between increasing the rate of pair generation to increase the key rate, or decreasing the rate of pair production to decrease the error rate \cite{Lim:08}. The total loss an entanglement-based QKD system can tolerate, and thus the longest transmission distance attainable, is limited by the rate of pair generation as well as detector characteristics and error rate.

The photon pair production rate of other experimental set-ups have optimized using numerical simulations  \cite{Lim:08,2008arXiv0809.0923R,PhysRevA.81.023835,2008arXiv0804.0891K}.  In other cases, a few different pump powers are attempted before settling on the observed power with the best visibility \cite{springerlink:10.1007/978-3-642-11731-2_14,1367-2630-11-6-065004,1367-2630-11-4-045013,PhysRevLett.94.150501}. Some experiments are limited by classical communication between detectors, and so the two-fold coincidence rate is optimized for the processing of detection events \cite{PhysRevLett.98.060503,2005OExpr..13..202R}. 

 Ma, Feng and Lo \cite{PhysRevA.76.012307} found a numeric solution for optimal squeezing parameter for QKD with entanglement using SPDC sources, but it requires root finding and produces negative values for certain channel efficiencies. Although the two-fold coincidence rate can be easily measured, the squeezing parameter is unmeasurable in practice, because the channel efficiency, dark count rate, and multi-order photon terms taint the measurement. We provide a practically useful, predictive model for the optimal two-fold coincidence rates. Such a model must operate with realistic bucket detectors and dark counts, and must require as input and produce as output only variables that are easily experimentally measureable. This model will help optimize QKD systems in real-time, and provides insights into the maximum possible distance and bit rate of entanglement-based QKD. We are able to produce this model by eschewing first-principles modeling in favor of a symbolic regression approach.

\section*{Background}
Quantum key distribution (QKD) uses the Heisenberg uncertainty principle to ensure secure key distribution protected from eavesdropping in an information theoretic secure manner\cite{paterson2004quantum}. 

QKD involves one party (Alice) sending quantum states (e.g. polarized photons) to a second party (Bob) \cite{bennett1984quantum}.The system's security depends upon the non-orthogonal complementary bases used to measure the quantum states. If Bob, or an eavesdropper, measures in a basis other than the one Alice used to prepare her state, his measurement will be random noise. If an eavesdropper measures in the wrong basis, they will introduce detectable errors in Bob’s measurement results, allowing Alice and Bob to abort key generation before any secrets are shared.

An example implementation of QKD involves generating polarization correlated entangled photon pairs with Alice and Bob measuring a photon from each pair in one of two random bases \cite{bennett1992quantum}. Alice and Bob measure a shared source of entangled photons in random bases. The key is formed from the results in which Alice and Bob measured in the same basis. In this paper, we call the coincidence rate the number of times per second where one of Alice's and one of Bob's detectors click within the same time window.

In the implementation described above, entangled photon states are created using two sources of pairs of polarization-correlated photons created by spontaneous parametric down-conversion (SPDC) in nonlinear material. These pairs can be made indistinguishable either by using a Sagnac interferometer loop \cite{shi2004generation}, by selecting  overlapping spatial modes in Type-II SPDC processes \cite{kwiat1995new}, or by stacking two nonlinear materials at orthogonal angles (known as sandwich sources) \cite{kwiat1999ultrabright}. 

While it is photon pairs that make the desired entangled states, SPDC  produces unwanted higher order photon states as well.  More than one photon can be detected by Alice or Bob during the same timing window. If two orthogonal detectors click, the result must be assigned to a random result for security reasons \cite{PhysRevA.81.052342}, as depicted in fig. \ref{fig:experiment_setup}. In QKD protocols, the more errors there are in a sifted key, the more of the sifted key needs to be revealed, and thus rendered useless, in order to perform error correction and privacy amplification.

\section*{Simulation}
Our simulation is written in Python using the QuTIP Quantum Toolbox in Python \cite{johansson2012qutip} \footnote{our code is avaliable at: http://qutip.blogspot.ca/2012/06/why-release-your-source-code-and.html}. We use a fock state representation following Jennewein et al. \cite{jenneweinjmo:11} and the mathematical description of SPDC and bucket detectors from Kok et al. \cite{RevModPhys.79.135}. The secure key rate is calculated using the QBER and two-fold coincidence probability in the inifinite key limit described in \cite{PhysRevA.76.012307}.

Scarani and Renner found that $1\e{6}$ raw bits must be exchanged in order to get a positive key \cite{scaranirenner08}. We define a system being usable if we have a secure key rate above 14~bits/s so that a positive key can be exchanged over the course of an hour.

Note that in order to calculate the secure key rate and two-fold coincidence rate per second, the secure key bit and two-fold coincidence probabilities must be divided by the experimentally defined coincidence window. The coincidence window is the maximum amount of time allowed to pass between Alice and Bob's detections in order for two detections to be considered a coincidence.
 
\section*{Results}

\begin{figure}
\centering
%this figure is clipped strangely 
\includegraphics[trim = 0mm 0mm 20mm 0mm, clip,width=\columnwidth]{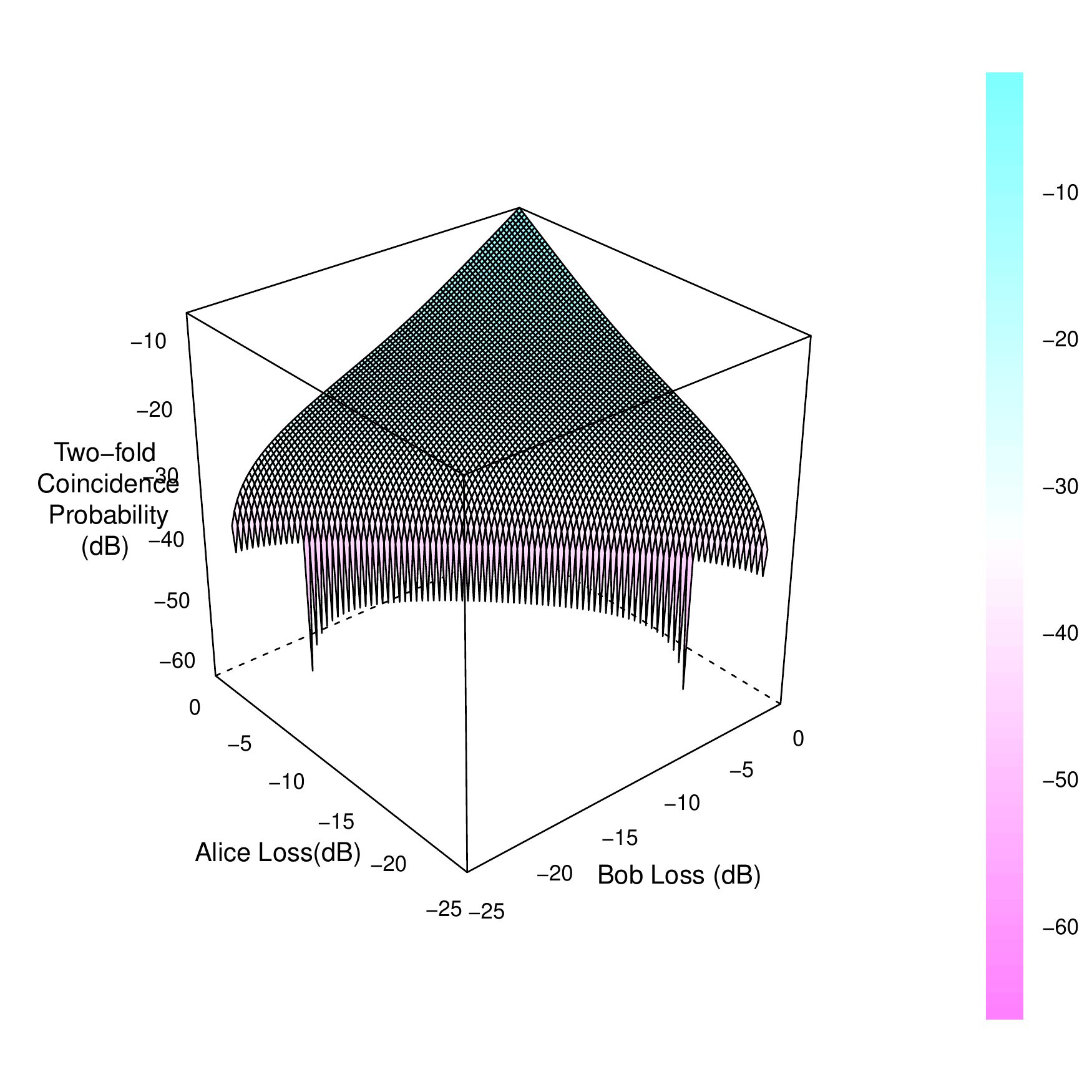}
\caption{Optimal two-fold coincidence probabilities predicted by the model based on channel loss and detector noise factor. In this case, $d_a = d_b=0$. Dark count rates are only a linear factor on the optimal two-fold coincidence rate.}
\label{fig:skr_graphs}
\end{figure}

\begin{figure*}
\centering
\includegraphics[trim = 0mm 0mm 10mm 0mm, clip, width=0.6\columnwidth]{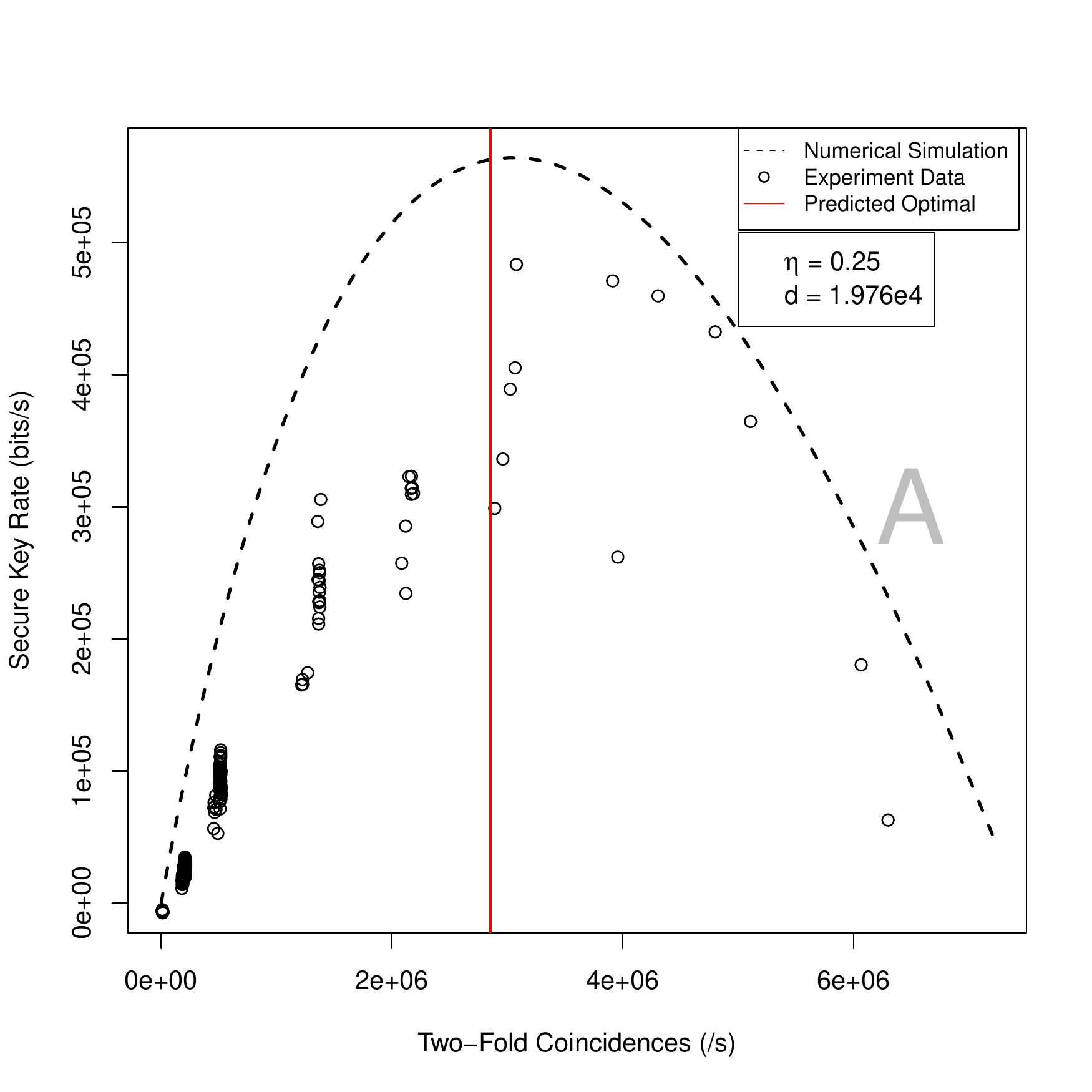}
\includegraphics[trim = 0mm 0mm 10mm 0mm, clip, width=0.6\columnwidth]{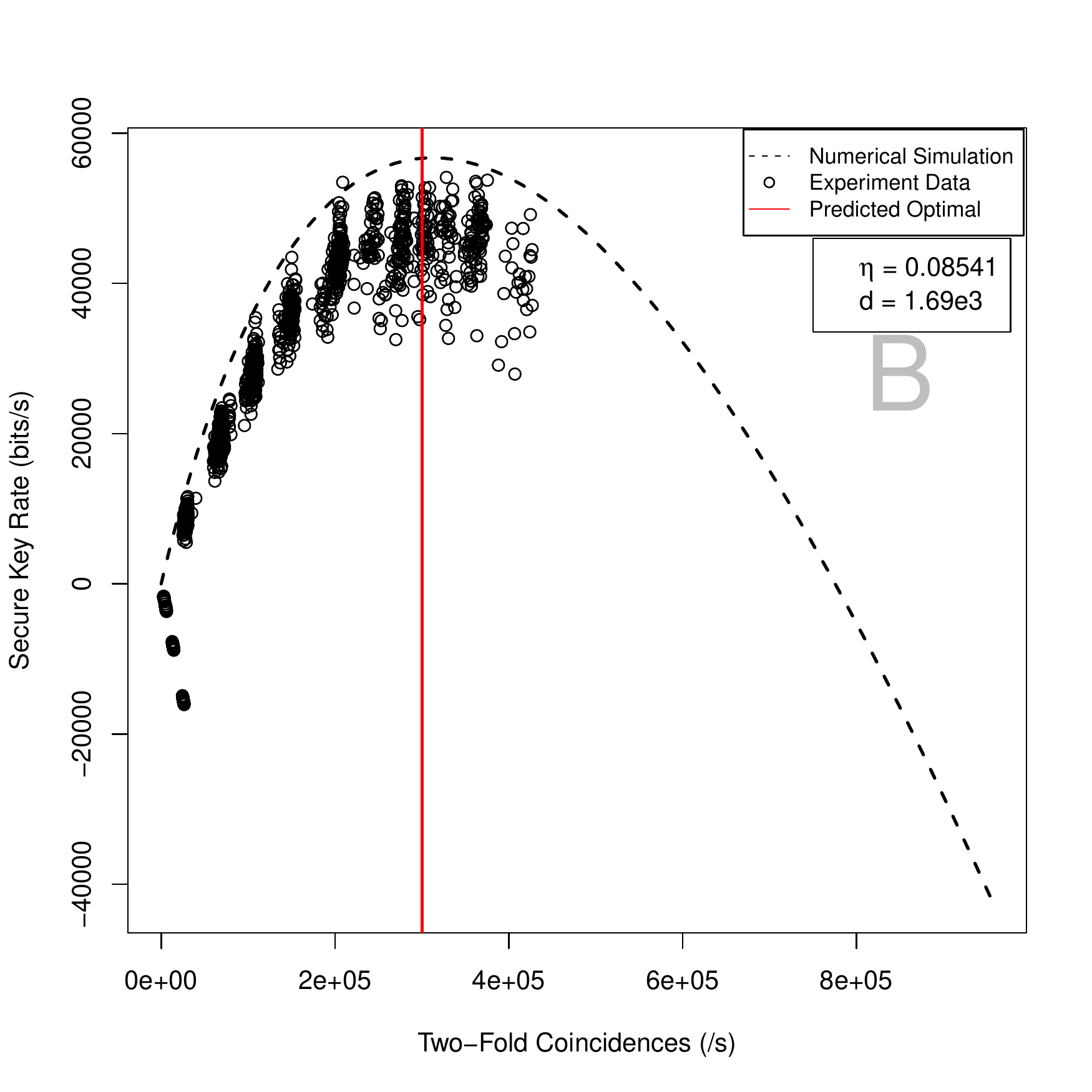}
\includegraphics[trim = 0mm 0mm 10mm 0mm, clip, width=0.6\columnwidth]{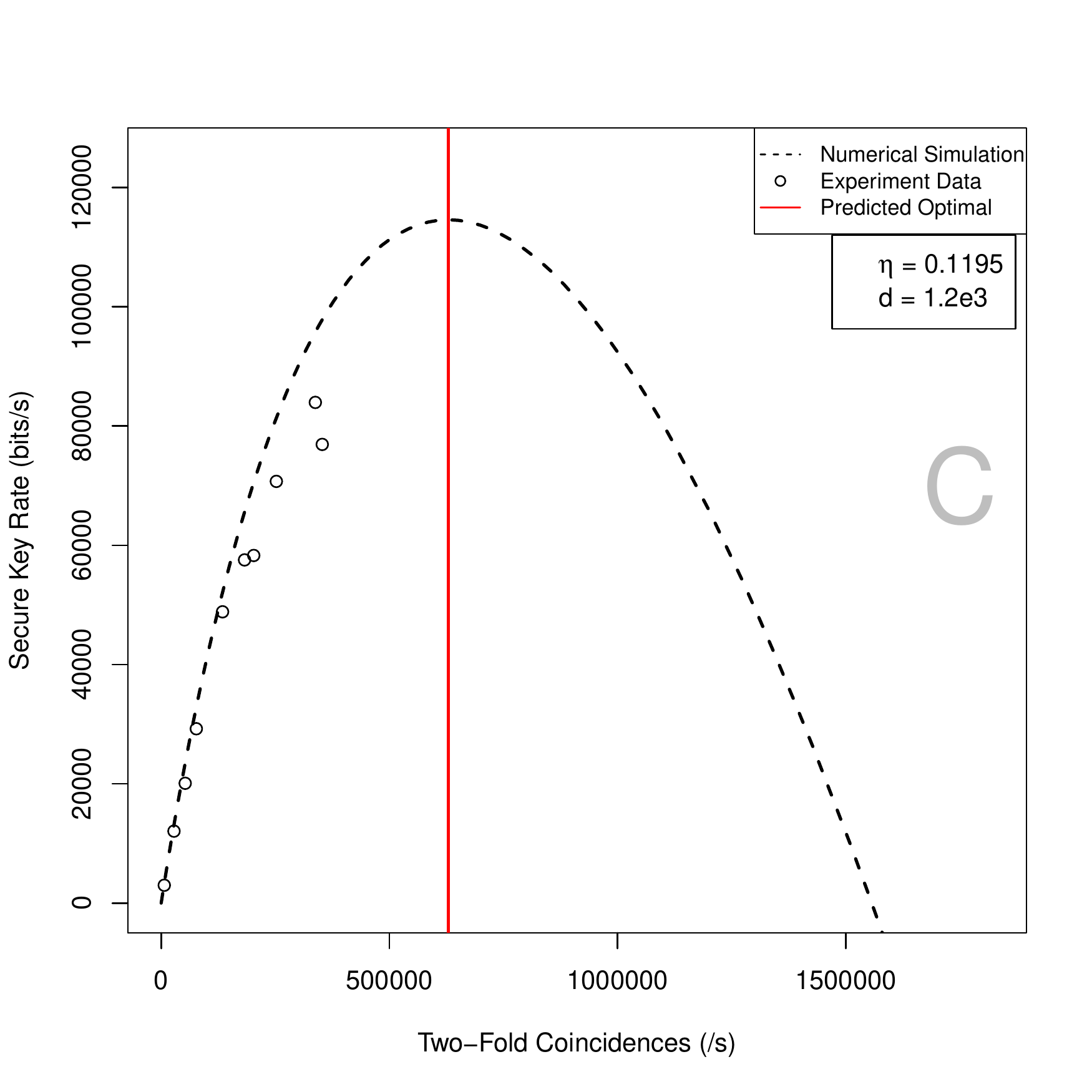}
\begin{tabular}{l P{3.5cm} P{3cm} P{3cm}}
\hline
Channel & Optimal from Experiment  &  Optimal from  Simulation	&  Optimal from Model \\
\hline
Noisy channel (A) &	2.79\e{5} $\pm$ 5\e{4} &3.01\e{6}    &	2.85\e{6} \\
Lossy channel (B) & 3.46\e{6} $\pm$  1\e{6} & 3.13\e{5} & 3.00\e{5}\\
Free-Space Channel (C)& -  &	6.29\e{5} &	6.29\e{5} \\
\hline
\end{tabular}
\caption{Estimation of SKR based on two-fold coincidence rates for different input power rates. Three different experimental conditions were used: uncovered polarization analyzer without neutral density filters ($d =1.976\e{4}$, $\eta= 0.25 \pm 1\e{-2}$, graph A), covered polarization analyzers with neutral density filters ($d = 1.69\e{3}$, $\eta = 0.085 \pm 1\e{-3}$, graph B), and after a free space channel ($d = 1.2\e{3}$, $\eta= 0.12 \pm 1\e{-2}$, graph C). The coincidence windows for A and C were 3.5~ns, the coincidence window for B was 2.5~ns.}

\label{fig:exp_data}
\end{figure*}

We found the squeezing parameter $\epsilon$ that maximized the secure key rate using Scipy Optimize. We used this optimal $\epsilon$ to calculate two-fold coincidence probabilities and secure key bit probabilities for 500,000 different combinations of the detector dark count probability per coincidence window, $d=\{0..0.1\}$  and the channel efficiency, $\eta = \{0..1\}$.  These values and free parameters were then provided to the symbolic regression tool \emph{Eureqa} \cite{Schmidt03042009}. Symbolic regression produces a predictive model from data comprized of arbitrary algebraic functions of the input. \emph{Eureqa} produced a model for two-fold coincidence probability that maximized the secure key rate:
\begin{multline}
\label{tf}
P_{tf} = A\sqrt{\eta_a\eta_b} +B(\sqrt{\eta_a\eta_b}^3\\
\sin{(C-D\sqrt{\eta_a\eta_b}-\eta_a-\eta_b)}-d_a-d_b)+E
\end{multline} 
where $A = 0.03579$ , $B = 0.23$ , $C = 1.162$ , $D = 2.496$ , and $E =-0.002444$ , and $\eta_a$, $\eta_b$, $d_a$, $d_b$ are the channel efficiencies and the background noises to Alice and Bob, respectively. In order to determine the two-fold coincidence rate that this corresponds to, this number must be divided by the coincidence window. This model is plotted in figure \ref{fig:skr_graphs} (for the case when $d_a = d_b = 0$).

\section*{Experimental Verification}
\begin{figure*}
\includegraphics[width=2\columnwidth]{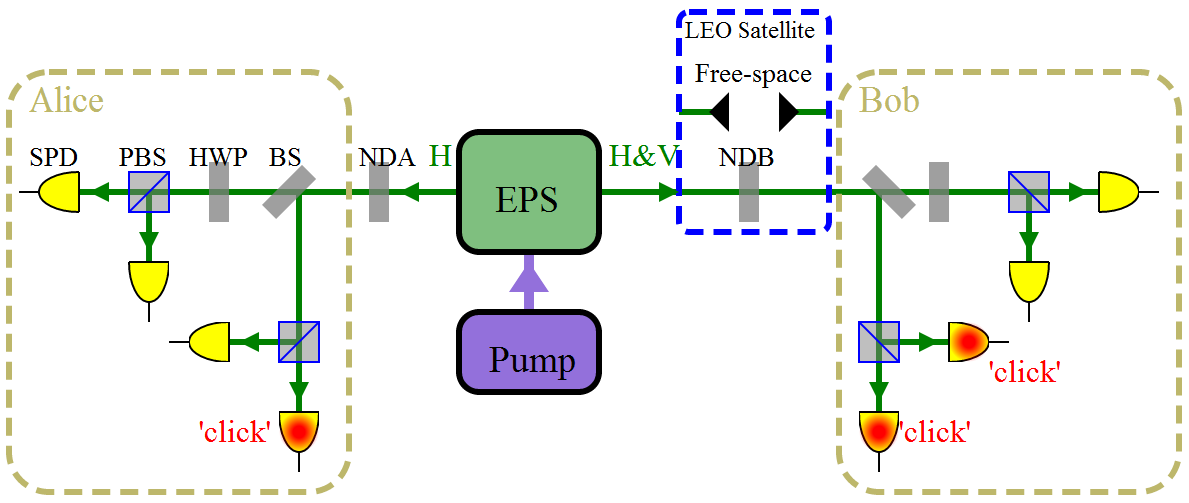}

\caption{While Alice received and correctly measured a horizontally polarized photon, Bob received a mult-photon state and measured two orthogonal detectors clicking at the same time, leading to an error. The entangled photon source (EPS) is a PPKTP crystal in a Sagnac configuration \cite{Fedrizzi:07}. The pump power was varied to observe the resulting coincidence rates and QBER. Loss was introduced by adding neutral density filters or a free-space channel. Noise was introduced by exposing Alice and Bob's detectors to flourescent light. In this diagram, HWP: half-wave plate, BS: beam splitter, NDA: neutral density filter, Alice, NDB: neutral density filter, Bob, BS: beam splitter, and PBS: polarising beam splitter.}
\label{fig:experiment_setup}
\end{figure*}
We verified the model by comparing the simulated data and the model to experimental data. We used a Sagnac source of entangled photons and passive basis choice polarization analyzers with Si-APDs for detection \cite{Erven:08}. Our experimental apparatus is demonstrated in fig. \ref{fig:experiment_setup}. We use passive measurements, while our simulation uses post-processing for active measurements, using passive measurements should be the same as active measurements if a post-processing scheme where measurements with conflicting results are discarded.

In typical operation, we used band-pass  filters to prevent the pump from entering the detector modules and to minimize errors due to dispersion. Typical operation also involved shielding the source from overhead light. Shielding the source reduced the dark count rate, as did introducing band-pass filters, although the latter also reduced channel efficiency.

In order to gather experimental data from a wide variety of dark count rates and channel efficiencies, we took experimental measurements under three conditions: covered polarization analyzers with neutral density filters ($d = 1.69\e{3}$, $\eta = 0.085 \pm 1\e{-3}$), uncovered polarization analyzer without neutral density filters ($d =1.976\e{4}$, $\eta= 0.25 \pm 1\e{-2}$), and after a free space channel ($d = 1.2\e{3}$, $\eta= 0.12 \pm 1\e{-2}$). The coincidence window for the first experiment was 2.5~ns, for the second two experiments it was 3.5~ns. In order to use the equation for optimal two-folds, the dark count rate must be scaled by the coincidence window.

Removing the band-pass filters reduced the channel loss but introduced far more dark counts. Adding neutral density filters partially occluded the source's output, decreasing channel efficiency. Removing the shielding of the Sagnac source increased the number of dark counts by a factor of 10. 

The pump power was varied between 0 and 50~mW in increments of 5~mW. The average coincidence rate and average QBER were then used to estimate the SKR using the asymptotic key rate. Results of these experiments are presented in fig. \ref{fig:exp_data}. The upper limit of the two-fold coincidences is limited by the maximum power of the pump laser. Fewer data points were collected on the noisy channel due to the limitation of the memory of the timetaggers not being able to handle the coincidence rate.

It is not possible to directly measure $\mu$ with existing experimental equipment, however, by comparing the empirically observed counts of detection coincidences and the calculated estimate of SKR to the corresponding values produced by the simulation, we can empirically validate the simulation.
\section*{Discussion}
From the experimental data and the numerical solution, we can estimate the optimal coincidence rate for the three channels, except for the free space channel where our laser power is insufficient to reach the coincidence rate. For the channels with experimentally determined optimal coincidence rates, the optimal determined from the numerical solution is within the margin of error for the experimental optimal. We now apply our model to investigate two QKD channels.

\subsection*{Application: Optimizing QKD with Satellites}

\label{appendix_satellite}
\begin{table*}
\centering

\begin{tabular}{|l|l l l l|}
\hline
& \multicolumn{3}{l}{\bf Ground Station Site} & \\
& Mountain (45~km) & Mountain (20~km) & Sea-Level (45~km) & Sea-Level (20~km) \\
\hline
\bf Total Key & & & &\\
\hline
Best Pass & 322692 &  301849 & 88216 & 97812\\
75\% Pass &183523 & 158257 & 44492 & 33010 \\
50\% Pass & 26058 &10294 & 2581 & -\\
\hline
\bf Additional Key & & & &\\
\hline
Best Pass & 2383 & 3651  & 444 & 1106 \\
75\% Pass & 4089  & 2676 & 507  & 1460 \\
50\% Pass  & 4681 & 6341 & 810 & - \\
\hline
\bf Percent Increase & & & &\\
\hline
Best Pass & 0.74\%&1.22\% & 0.46\% & 1.27\% \\
75\% Pass & 2.65\% &1.48\% & 1.15\% & 4.63\% \\
50\% Pass & 21.90\% & 62.34\% & 45.75\%  & - \\
\hline
\end{tabular}

\caption{Estimated additional key and total key generated by optimized variable two-fold coincidence rate compared with a fixed coincidence rate. The conditions are: source and transmitter on a mountain, or at sea-level, source and transmitter 20~km away from a city, source and transmitter 45~km away from Ottawa. There is no data for the median pass for a transmitter at sea-level and 20~km from Ottawa because it is not possible to exchange a key under these conditions. The median passes show the most improvement}
\label{additional_key}
\end{table*}

\begin{figure*}
\centering

\includegraphics[trim = 0mm 0mm 0mm 0mm, clip, width=\columnwidth]{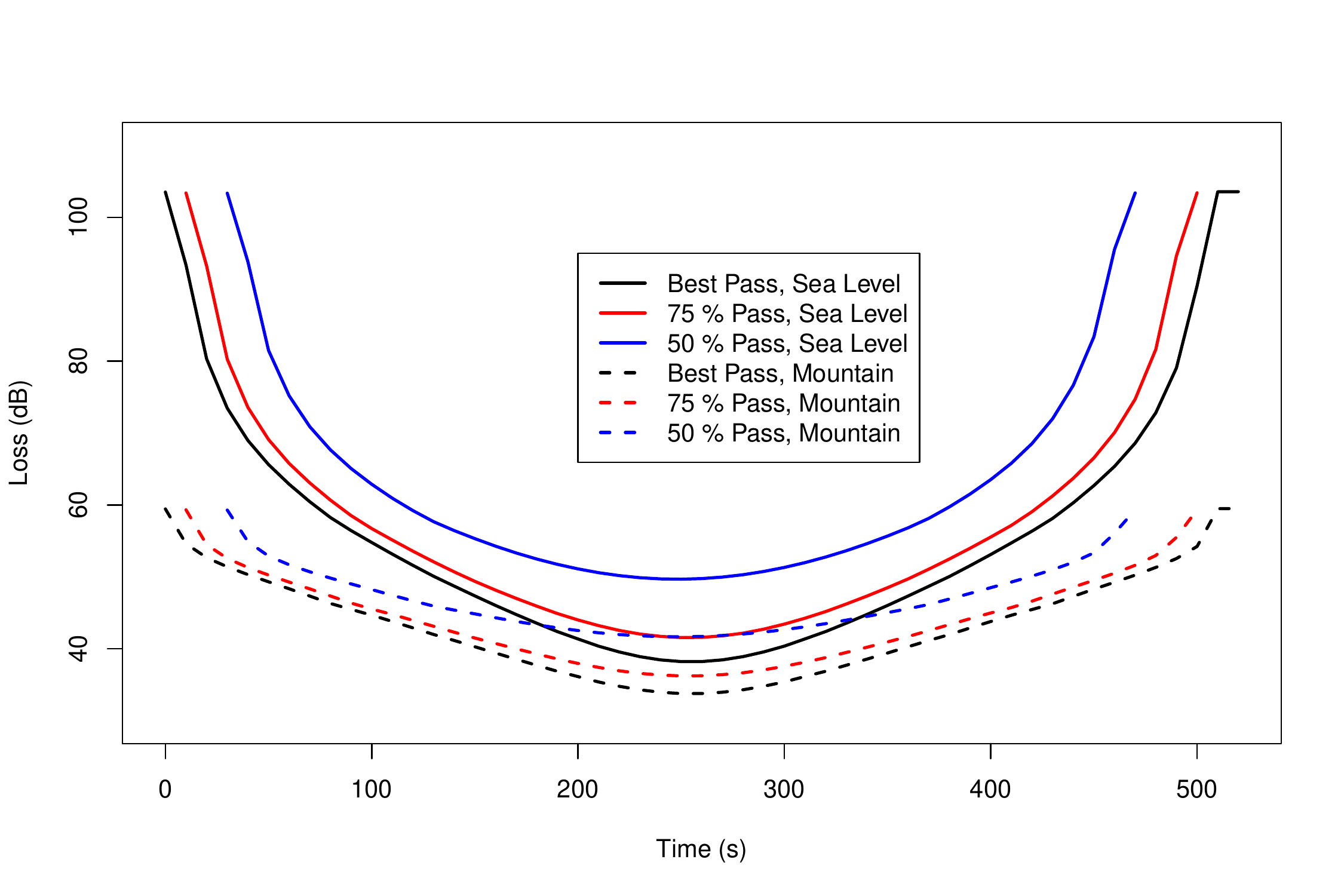}
\includegraphics[trim = 0mm 0mm 0mm 0mm, clip,width=\columnwidth]{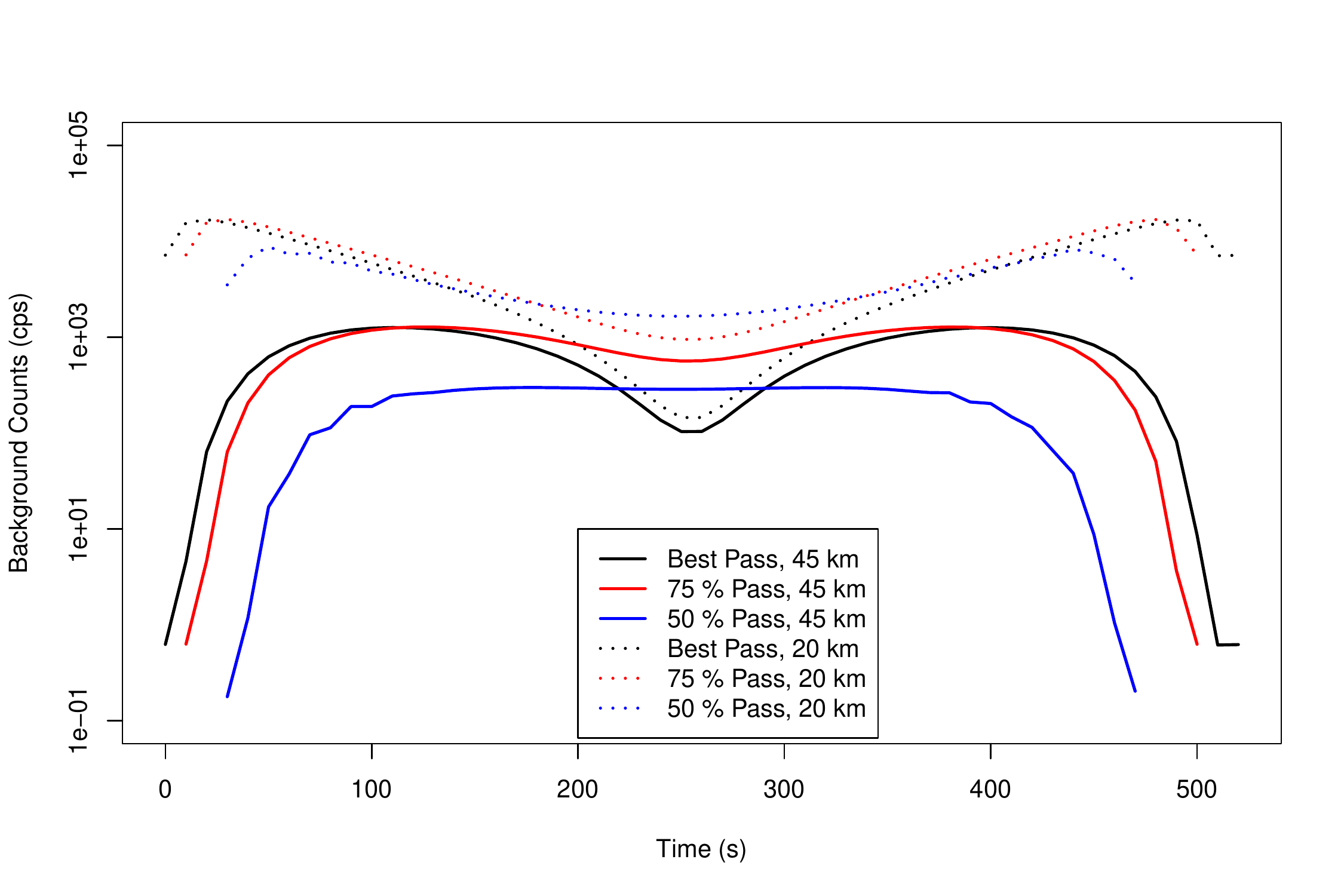}

\caption{Simulated channel loss and background noise for the satellite simulations used in this paper. The graph on the left displays the amount of loss from transmission through the atmosphere. Solid lines are conditions for a ground receiver in the country, at sea-level. Dashed lines represent conditions for a ground receiver in the country, on a mountain, and dotted lines represent conditions at sea-level closer to the city. Height has little effect on the background noise, and proximity to a city has little effect on the loss.}
\label{satellite_conditions}
\end{figure*}

\begin{figure*}
\centering

\includegraphics[trim = 0mm 0mm 0mm 0mm, clip, width=\columnwidth]{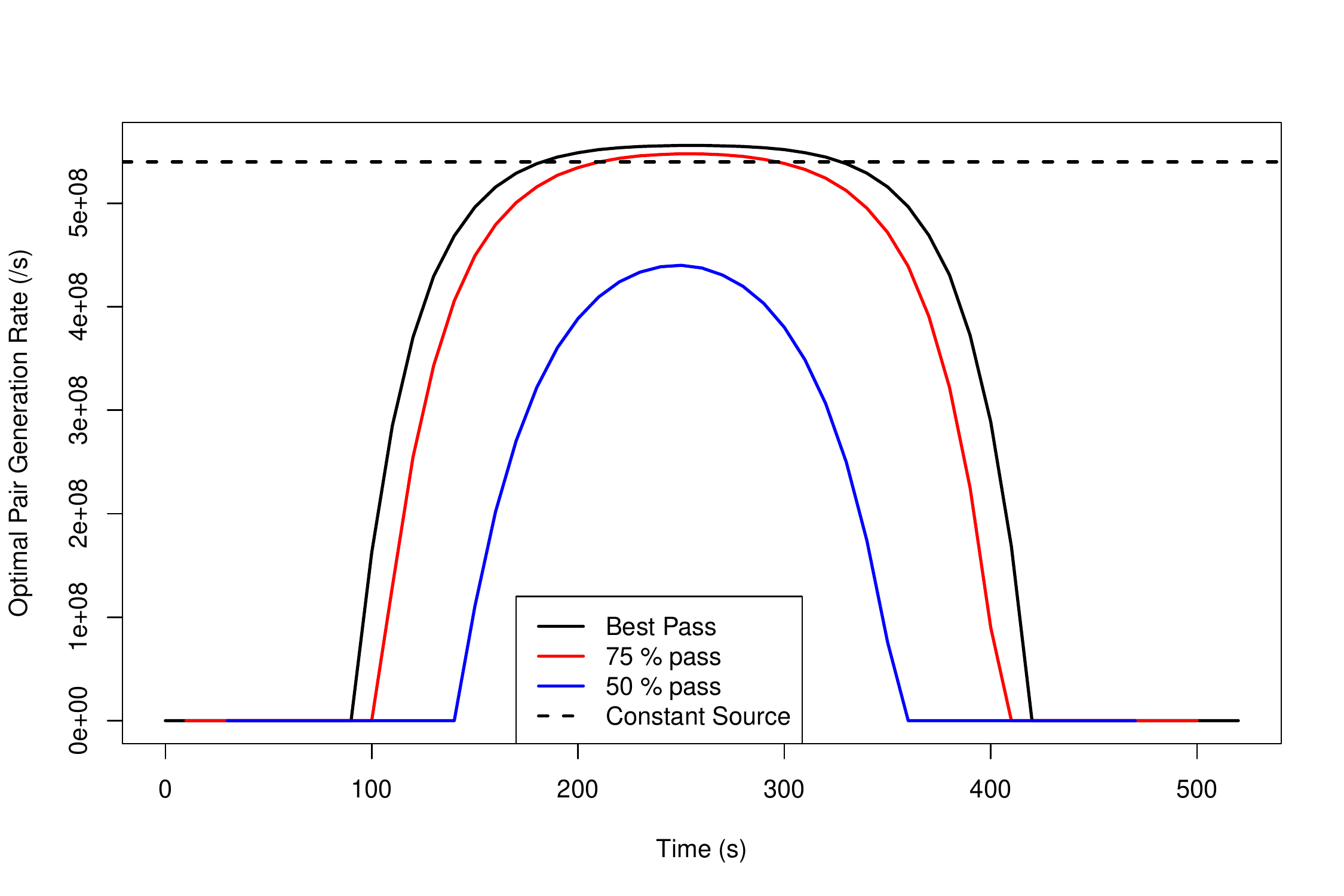}
\includegraphics[trim = 0mm 0mm 0mm 0mm, clip,width=\columnwidth]{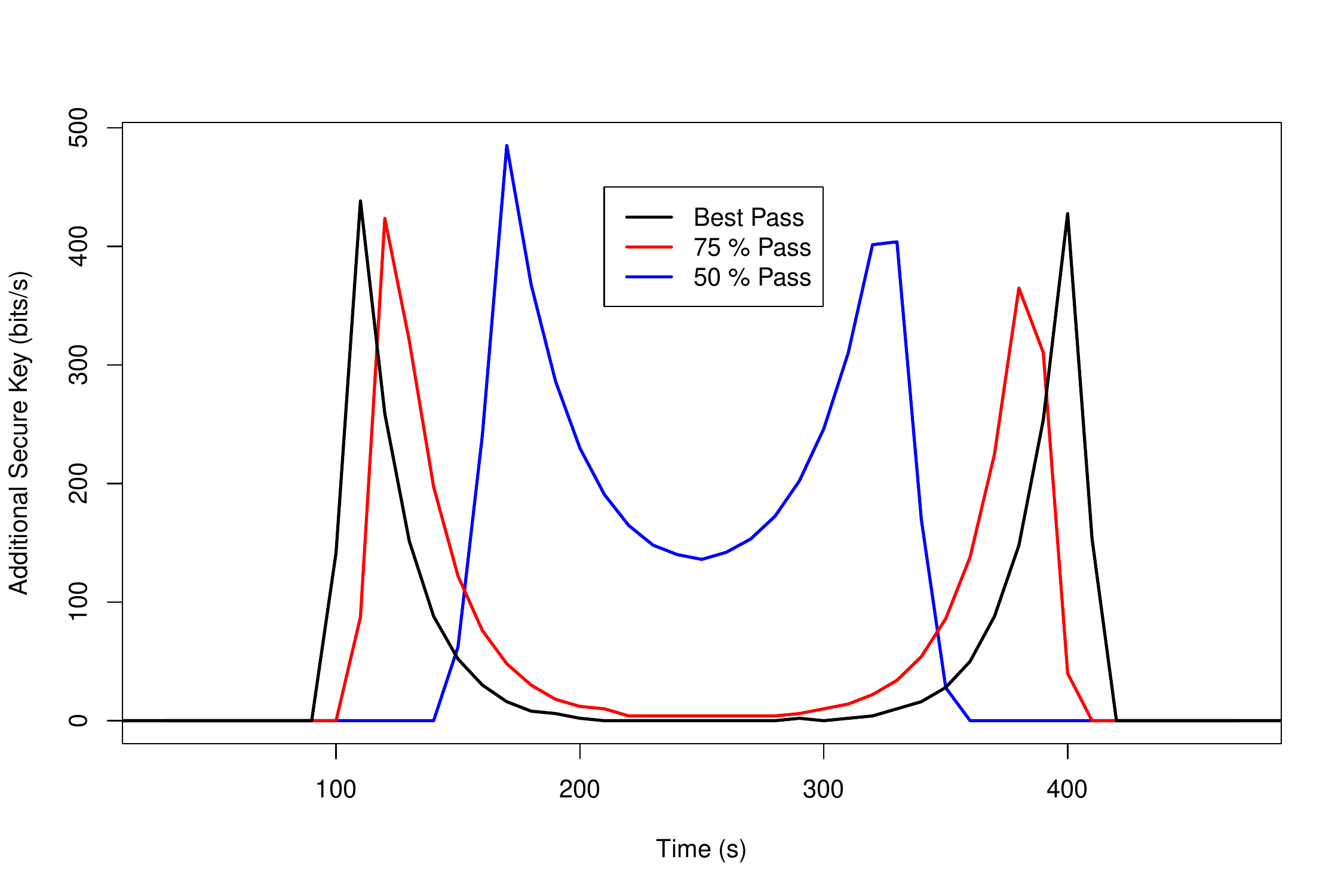}

\caption{Optimal pair generation rate and additional key from simulated satellite passes. In the left graph, pair generation rate is measured by unit bucket detectors at the source, and the dashed line represents the constant value for pair generation rate used for comparison. The graph on the right shows the additional secure key rate with the variable optimized pair generation rate over the constant optimized pair generation rate.}
\label{satellite_models}
\end{figure*}

We simulate loss and detector dark counts in a satellite uplink scenario \cite{JP2012inpreparation}, meaning a source on  the ground with one photon going to a LEO satellite while the other is measured on the ground, over a year of continuous usage. The ground stations are located on mountains (2.4~km above the ground) and 45~km outside of Ottawa, Canada. The ground telescopes had an aperture of 25~cm and the satellite telescope had an aperture of 20~cm. A low-earth-orbit satellite has a period of 1.6 hours, and the values of loss and dark counts are constantly changing as the satellite passes from horizon to horizon in the transmitter's field of view.

Loss and detector dark counts are used to calculate the optimal two-fold coincidence rates. We then estimate the SKR using this optimal two-fold coincidence rate, given detectors with a quantum efficiency of 50\% and a dark count rate of 100~c/s  (figure  \ref{satellite_models}). We also calculate the pair generation rate by using bucket detectors of unit efficiency at the source. Passes are ordered by the length of visible time between the transmitter and the satellite. We compare this against the estimated SKR from the pair generation rate fixed at a power that maximizes the secure key rate over all passes. We assume a coincidence window of 0.5~ns.

Adjusting the two-fold coincidence rate dynamically does not improve the key rate for good satellite passes (an increase of $0.75\%$ for the best link, and 2.65$\%$ during the course of the 75th percentile link, for the transmitter on 45~km from Ottawa on a mountain, the best pass both in terms of background counts and loss). The total key, additional key and percent increase for a variety of simulations and passes are presented in table \ref{additional_key}. Adjusting the source rate gives the biggest increase for passes where the loss is worst or the background counts increase. On the best passes, the usable time of the satellite pass increases by 10~s over the course of a 250~s pass. On median pass, the usable time more than doubles. This means that with optimization, many more satellite passes which were previously infeasible due to the high loss are now usable.

\subsection*{Application: Optimizing QKD with Fiber Optics}

\label{appendix_photondetection}

\begin{sidewaystable*}[p]
\centering
\vspace{9cm}
\begin{tabular}{p{3cm} p{1.74cm} p{1.78cm} p{1.74cm} p{1.75cm} p{2.45cm} p{2.45cm} p{1.8cm}  p{2.45cm} p{2.45cm} p{1.8cm}}
\hline
%\vspace{9cm}
Type & Wave-length (nm) & Time Resolution (ps) & Effi-ciency (\%) & Dark Counts (per s) & \multicolumn{3}{c}{Key in an Hour} & \multicolumn{3}{c}{Asymptotic Time Key} \\
& & & & & Optimal Coincidence Rate (After Channel) (pairs/s) & Optimal Pair Generation Rate & Loss Budget (dB) & Optimal Coincidence Rate (After Channel) (pairs/s) &  Optimal Pair Generation Rate & Loss Budget (dB) \\
\hline
Visible Range  & Detectors & & & & & & & & &  \\
\hline
Si-SPAD \\(thin junction) & 550 & 35 & 52 & 300 & 857 & 9.65\e{7} & 30.6 & 17.9 & 9.45\e{7} & 39.0 \\
Quantum Dot & 550 & 1.5\e{5} & 13 & 0 & 768 & 1.51\e{3} & 6.54 & 0.0112 & 1.46\e{3} & 30.7 \\
Si-SPAD \\(thin junction) & 850 & 35 & 12 & 300 & 857 & 5.77\e{6} & 24.2 & 17.9 & 5.64\e{6} & 32.6 \\
Si-SPAD \\(thick junction) & 830 & 500 & 45 & 25 & 853 & 5.17\e{6} & 24.2 & 1.25 & 5.02\e{6} & 38.3 \\
\hline
Telecom Range  & Detectors & & & & & & & & &  \\
\hline
InGaAs SPAD & 1550 & 500 & 3 & 4\e{4} & 8660 & 1.24\e{4} & 6.01 & 7160 & 1.15\e{4} & 6.30 \\
InGaAs SPAD & 1550 & 55 & 11 & 2,925 & 1010 & 2.92\e{6} & 22.4 & 7.70 &1.83\e{6} & 32.2 \\
Tungsten transition edge sensor  & 1550 & 9\e{4} & 88 & 10 & 778 & 1.01\e{5} & 16.1 & 0.124 & 6.15\e{4} & 34.1 \\
NbN nanowire & 1557 & 60 & 1 & 10 & 958 & 1.19\e{4} & 10.5 & 10.7& 1.19\e{4} & 20.3 \\
Up-conversion assisted hybrid photodetector  & 1550 & 200 & 4 & 3\e{4} & 3000 & 6.98\e{4} & 12.1 & 1650 & 5.74\e{4} & 13.0 \\
Up-Conversion  & 1560 & 50 & 5 & 5\e{4} & 2540 & 4.56\e{5} & 16.5 & 1190 & 3.62\e{5} &  17.8 \\

\hline
\end{tabular}
\caption{Optimal coincidence rates, and the Loss Budget in order to get a key (50,000 secure key bits) in an hour or ten years for symmetric entanglement-based QKD for common single photon detectors for telecom frequency photons. Table adapted from Buller and Collins \cite{buller2010single}.}
\label{detector_table}
\end{sidewaystable*}

We use the values reported for time resolution, detector efficiency, and detector dark counts for several types of detectors  \cite{buller2010single} to compute loss budgets. The loss budget is the largest loss, given an optimal pumping rate, for which it is possible to transmit a secure key of at least 50,000 bits \cite{scaranirenner08} in a given time period. For each detector, we compute loss budgets for periods of one hour and the asymptotic limit of time, including loss from imperfect detectors. To obtain the maximum loss permissible for a channel using these detectors, we then subtract the detector ineffiencies from \cite{buller2010single} off of the loss budget. Our findings are summarized in table \ref{detector_table}.

Our estimates should be taken with a caveat that they do not account for finite size effects, which increase the QBER at very low coincidence rates \cite{scaranirenner08}.

\subsubsection*{Dark Optical Fiber Implementations}

The largest loss budget for a key in an hour from table \ref{detector_table} is 22.4~dB at 1550~nm. If we assume a continuous single mode fiber channel link, no loss from other sources (such as insertion loss), a fiber loss of 0.17~dB/km \cite{corningsheet}, and symmetric links \cite{scheidl2009feasibility}, the maximum possible distance predicted for entanglement-based QKD systems in fiber optic cables is 263.5~km. This number is calculated by dividing the loss budget by the loss per kilometer and multiplying by two for the symmetric links. In the asymptotic limit, this loss budget goes up to 34.1~dB or 401.2~km.

Visible wavelengths suffer from much higher losses in fiber, from 3~dB/km at 800~nm \cite{meyer2010quantum}, to 30~dB/km at 515~nm \cite{thorlabss405}. Using the calculation above, this means that the furthest a visible-wavelength implementation could travel in fiber is 16.3~km for a key in an hour and 25.5~km for a key in the asymptotic limit at 800~nm. Thus, although visible light detectors have greater detection efficiency and fewer dark counts, they are less useful for long-distance fiber implementations due to the attenuation of visible light in fiber. 

\subsubsection*{Bright Optical Fiber Implementations}

QKD with entanglement distribution has been implemented on bright (carrying classical data) standard single-mode telecommunications fibers. This can be done by sending the quantum information at an unused wavelength in dense wavelength devision multiplexing protocols \cite{chapuran2009optical,1367-2630-12-6-063027,1367-2630-11-4-045012}, or by using a wavelength in the visible range, far from infrared telecommunications wavelengths \cite{Holloway:11}. Using multiple close wavelengths on the same fiber leads to wave-mixing processes, such as stimulated brillouin scattering and four-wave mixing \cite{agrawal2001nonlinear}. Four-wave mixing processes are a concern for experimentalists of telecommunications systems, but are an obstacle to QKD systems. QKD systems operate at much lower optical powers than the classical communications traffic (0.1-10~pW compared to 0.1-100~mW), so it is more likely for the classical channels to mix and spread into the quantum channels than vice-versa. The photons produced by wave-mixing processes are generated in random bases, and can be interpreted as detector dark counts in analysis.

In simulation, we can determine the maximum `noise budget' - meaning the maximum dark count probability that can be tolerated given a channel efficiency. We find that this noise budget approximately follows a rational equation, where in order to get a positive key, the maximum tolerable dark count probability is:
\begin{equation}
\label{noise_budget}
d \leq \frac{0.0732\eta_a\eta_b}{\eta_a + \eta_b}
\end{equation}
where $\eta_a$ and $\eta_b$ are the channel efficiencies to Alice and Bob, respectively. An experiment could be optimized by following this limit.

For implementations with visible wavelengths, the impact of mixing and scattering processes is negligible due to the wavelength distance between classical and quantum signals. Therefore, systems on bright fibers with visible wavelengths have the same maximum distance as visibile wavelength implementations on dark fibers (~16~km).

\section*{Conclusions}

We have used realistic detector models with correct treatment of double pairs to determine the two-fold coincidence probability that would be measured in a given entanglement-based QKD system when the system has the largest secure key bit probability. We have used symbolic regression to create an equation relating the optimal two-fold coincidence probability to the detector dark count probability and the system loss. We have also taken experimental data to show that our simulation matches reality and that our model accurately indicates the maximum under extreme experimental conditions..

We hope that in finding this relation, we have provided future experimentalists with a useful tool. At the moment many demonstrations of QKD with entangled photon pairs rely on low numbers of coincidences where the visibility is high \cite{Erven:08,Holloway:11}. However, as detectors and sources improve and experimentalists compete for the new distance record, the issue of the tradeoffs between coincidence rate and visibility will have to be adressed. Our model provides a simple method for maximizing the throughput of QKD systems, which relies only on presently measurable variables. We believe this model will allow for near real-time optimisation in pump power in real-world implementations such as on active telecommunications networks and satellite transmission, where background and losses change quickly and unpredictably. It could also provide a starting point for future theoretical exploration of this phenomenon.

\section*{Acknowledgements}

The authors would like to acknowledge Xiongfeng Ma for sharing his data. We also gratefully acknowledge support for this work from NSERC, CSA, CIFAR, CFI, the Ontario Ministry of Training, Colleges and Universities, the David R. Cheriton school of Computer Science. 

\bibliographystyle{ieeetr}	% (uses file "plain.bst")
\bibliography{entanglement,myfootnotes}

\section{Comparison between our optimal $\mu$ and Ma's optimal $\mu$}
\label{appendix_mu}
\begin{figure}

\includegraphics[width=8cm]{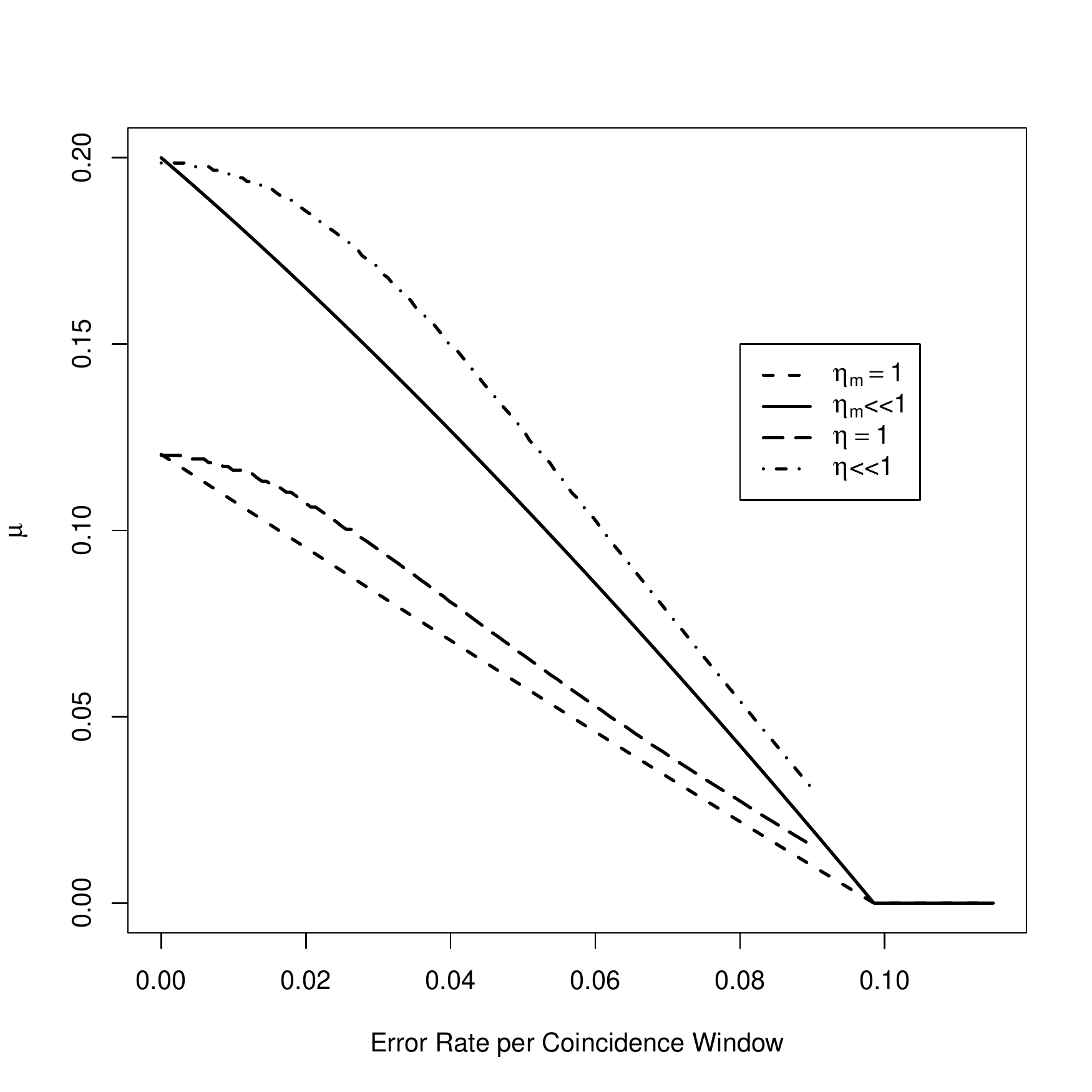}
\caption{Comparison of our model for the optimal $\mu$ to the numerical solution to $\mu$ found by Ma, Fung and Lo \cite{PhysRevA.76.012307}. $\eta_m$ are the channels used by Ma, Fung and Lo. }
\label{comparison_models}
\end{figure}

In the appendix of Ma, Fung and Lo's paper on QKD with entangled photon sources, they use their calculations for photon gain, secure key rate in the asymptotic limit of shared bits, and a model of loss and error rates in order to determine the optimal value for the photon pair production rate, $\mu$. They simplify for two cases, for a lossless channel $\eta = 1$ and a very lossy channel $\eta << 1$. They come up with an approximate relation which must be numerically solved in order to determine $\mu$ in terms of the intrinsic detector error. 

Our simulation differs from Ma's equations in three ways. We use detector models with poissonian distributions of dark counts. We do post-processing to assign double clicks to random bases, and we look at detector dark counts, not detector error, which flips the state of some incoming photons instead of adding noise to the detection probability. Although our model determines the optimal measured two-fold coincidence rate, we calculate the theoretical $\mu$ at the same time for various detector dark counts and loss. Our definitions of error are slightly different but we believe that they are similar enough to allow for direct comparison between our model and theirs, which we do in figure \ref{comparison_models}.

\end{document}